# Comparative Evaluation of PyTorch, JAX, SciPy, and Neal for Solving QUBO Problems at Scale

Pei-Kun Yang

E-mail: peikun@isu.edu.tw

ORCID: https://orcid.org/0000-0003-1840-6204

**Abstract**

Quadratic Unconstrained Binary Optimization (QUBO) is a versatile framework for modeling combinatorial optimization problems. This study benchmarks five software-based QUBO solvers: Neal, PyTorch (CPU), PyTorch (GPU), JAX, and SciPy, on randomly generated QUBO matrices ranging from 1,000×1,000 to 45,000×45,000, under six convergence thresholds from $10^{-1}$ to $10^{-6}$. We evaluate their performance regarding solution quality (energy) and computational time. Among the solvers tested, Neal achieved the lowest energy values but was limited to problems with up to 6,000 variables due to high memory consumption. PyTorch produced slightly higher energy results than Neal but demonstrated superior scalability, successfully solving instances with up to 45,000 variables. Its support for GPU acceleration and CPU multi-threading also resulted in significantly shorter runtimes. JAX yielded energy values slightly above those of PyTorch and was limited to 25,000 variables, with runtimes comparable to PyTorch using GPU. SciPy was the most constrained solver, capable of handling only up to 6,000 variables. It consistently produced the highest energy values and required the longest computation times. These findings highlight the trade-offs between solution quality, scalability, and runtime efficiency, and suggest that PyTorch is the most balanced choice for large-scale QUBO problems when computational resources permit.

**Introduction**

Quadratic Unconstrained Binary Optimization (QUBO) problems constitute the mathematical foundation for a wide range of combinatorial optimization tasks, where the goal is to find a binary vector that minimizes the quadratic form $F(X) = X^T Q X$, with $Q$ being a real symmetric matrix [1]. QUBO problems have broad applicability in domains such as portfolio optimization, job scheduling, feature selection, protein folding, and protein–ligand binding [2-8]. From the perspective of statistical thermodynamics, many of these problems reduce to searching for the global minimum in a complex and often rugged energy landscape. While one-dimensional optimization problems can be solved analytically or via simple iterative techniques such as the Newton–Raphson method [9], the difficulty of finding global optima increases exponentially with dimensionality, making high-dimensional QUBO problems computationally intractable using brute-force approaches.

To address this challenge, both classical and quantum-inspired methods have been proposed. Quantum hardware platforms such as those developed by D-Wave [10,11] and digital annealers introduced by Fujitsu [12,13] represent promising architectures that exploit hardware acceleration to solve QUBO models. In contrast, classical heuristics developed over decades have continued to evolve through techniques such as simulated annealing [1,2,14-17]. While classical methods remain widely used, emerging quantum and quantum-inspired architectures are expected to provide faster convergence to better minima in high-dimensional landscapes [18].

Nevertheless, access to commercial quantum hardware remains limited due to high operational costs, strict hardware constraints, and proprietary restrictions. Consequently, recent attention has shifted to software-based solvers inspired by quantum principles but implemented on classical hardware. These include simulated annealing frameworks, gradient-based optimizers, and physics-informed dynamical systems.

This study focuses on five software-based QUBO solvers: Neal, PyTorch (CPU and GPU), JAX, and SciPy. All solvers are implemented as open-source Python libraries [19-21], making them accessible to various researchers. Neal, developed by D-Wave Systems, is a simulated annealing sampler often used as a classical baseline for evaluating quantum-inspired methods. PyTorch, initially built for deep learning, supports automatic differentiation and GPU acceleration, making it suitable for solving relaxed QUBO formulations via gradient descent. JAX and SciPy, though designed for general-purpose numerical computation, can be adapted to solve QUBO problems using differentiable relaxation and bounded optimization.

We introduce a continuous vector **x**, which is projected to the interval [0,1] using a sigmoid function to approximate binary constraints in gradient-based solvers. A subsequent application of the Heaviside step function produces a binary vector suitable for evaluating the original QUBO energy. This transformation allows the continuous optimization of a relaxed objective function using gradient-based methods.

To assess the practical performance of these frameworks, we construct a systematic benchmarking pipeline to evaluate all five solvers across multiple QUBO instance sizes. Our evaluation criteria include runtime, solution quality, and scalability. Due to the high cost of commercial quantum devices such as D-Wave and Fujitsu's digital annealing systems and the limited availability of proprietary GPU-accelerated solvers, this study emphasizes freely available and open-source solutions. Through this comparative analysis, we aim to provide researchers and practitioners with practical guidance for selecting appropriate solvers based on problem scale, desired accuracy, and available computational resources.

**Methods**

**Generating Symmetric QUBO Matrices for Benchmarking.** To generate benchmark datasets for large-scale QUBO problems, we designed a Python-based procedure that constructs symmetric matrices $Q \in \mathbb{R}^{n \times n}$. Each matrix element is initialized with a random value uniformly drawn from the interval [−5, 5] using PyTorch. The matrix is averaged with its transpose to ensure the symmetry required by the QUBO formulation. In our experiments, we generated QUBO instances with four different sizes, specifically $n$ = 1,000, 6,000, 25,000, and 45,000, to evaluate solver performance under varying levels of computational complexity.

**Implementation of PyQUBO-Based Solver with Neal.** To evaluate the effectiveness

of simulated annealing on QUBO problems, we implemented a PyQUBO-based solver that interfaces with the Neal backend, a classical simulated annealing sampler provided by D-Wave's Ocean SDK. The compiled QUBO model is optimized using Neal's SimulatedAnnealingSampler, which performs annealing based on a geometric beta schedule ranging from $\beta_{min} = 0.1$ to $\beta_{max} = 4.0$, with the number of reads set to 10.

For each solver configuration and QUBO instance size, the evaluation is repeated five times using different initial random seeds to account for stochastic variability in the sampling process. The solution with the lowest energy is selected for reporting among the returned samples.

We record the wall-clock time required to compile and solve each instance to quantify performance. The final solution energy is taken from Neal's returned energy value. The binary solution vector, corresponding energy, and runtime are saved for analysis. All simulations are executed on the CPU, and PyTorch tensors are used throughout to facilitate efficient matrix operations and integration with the broader evaluation pipeline. Due to hardware memory constraints, the Neal-based solver only applies to QUBO instances with variable sizes of $n = 1,000$ and $n = 6,000$.

**Implementation of PyTorch-Based QUBO Solver (CPU and GPU).** To solve large-scale QUBO problems using gradient-based optimization, we implemented a PyTorch-based solver that approximates binary variables through differentiable relaxation. A real-valued parameter vector **x** is optimized by gradient descent and projected to a relaxed binary form **x'** using the sigmoid function:

$$\mathbf{x'} = \sigma(s(\mathbf{x} - 0.5)) \quad (1)$$

where $\sigma$ denotes the sigmoid function and $s$ is a slope coefficient controlling the sharpness of the projection. To obtain a final binary solution, a hard threshold is applied through the Heaviside step function:

$$\mathbf{x''} = u(\mathbf{x'} - 0.5) \quad (2)$$

This transformation enables continuous optimization of a discrete objective, accelerating convergence for modern optimizers such as Adam and a ReduceLROnPlateau learning rate scheduler.

Both CPU and GPU versions were implemented using PyTorch. To determine convergence, we introduced a moving average window over recent loss values and applied a dynamic early stopping criterion based on the relative change in this average. Optimization terminates when the change falls below a predefined threshold or when a maximum patience value is exceeded.

To investigate the impact of convergence precision, we tested six different threshold values: $10^{-1}$, $10^{-2}$, $10^{-3}$, $10^{-4}$, $10^{-5}$, and $10^{-6}$. Each configuration was evaluated using four problem sizes: $n = 1,000$, $6,000$, $25,000$, and $45,000$. The maximum number of optimization steps was set to 1,000,000 for all runs. However, training could terminate early if the relative change in average loss over a moving window fell below the specified threshold. Each configuration was repeated five times using different initial random seeds. At the end of each run, the final binary solution was reconstructed, and its energy was computed using only the upper triangular portion of the **Q** matrix to avoid redundant calculations.

**Implementation of JAX-Based QUBO Solver.** To solve large-scale QUBO problems through continuous relaxation and gradient-based optimization, we implemented a solver using JAX and the Optax optimization library. The binary constraint was relaxed using a sigmoid-based projection (Equation 1), allowing the optimization process to operate in a continuous space. The resulting relaxed loss function is differentiable and supports efficient gradient computation through jax.grad.

We employed the AdamW optimizer from the Optax library, configured with a learning rate of 0.01 and a weight decay of $10^{-5}$. After each update step, the parameters were clamped within the range [−5,5] to maintain numerical stability. Optimization proceeded for 1,000,000 steps, but early stopping was enabled using a moving average of the loss. Specifically, training was terminated when the relative change in average loss over a defined window dropped below a specified threshold or when the patience limit was exceeded.

To investigate the effect of convergence criteria on performance, we experimented with six different threshold values: $10^{-1}$, $10^{-2}$, $10^{-3}$, $10^{-4}$, $10^{-5}$, and $10^{-6}$. After training, the soft solution was converted to a binary vector using a step function (Equation 2). This JAX-based solver was evaluated on QUBO instances of size $n$ = 1,000, 6,000, and 25,000. Each configuration was executed five times using different initial random seeds to account for variability in optimization dynamics.

**Implementation of SciPy-Based QUBO Solver.** We implemented a continuous relaxation approach using the L-BFGS-B algorithm provided by the SciPy optimization library. To enable differentiable optimization, the binary constraint was relaxed using a sigmoid-based projection (Equation 1). The resulting continuous objective was minimized using SciPy's minimize() function with the L-BFGS-B method. The initial solution was sampled from a standard normal distribution, and each parameter was constrained within the bounds [−5,5] to maintain numerical stability.

Optimization was performed with a maximum iteration limit of 1,000,000. Termination was also controlled by a convergence threshold, which was varied across six levels: $10^{-1}$, $10^{-2}$, $10^{-3}$, $10^{-4}$, $10^{-5}$, and $10^{-6}$, to examine the influence of convergence precision on both runtime and solution quality.

The continuous solution was binarized using a step function (Equation 2) after termination. The energy of the resulting binary solution was computed using only the upper-triangular portion of the $Q$ matrix to avoid redundant calculations due to symmetry. Experiments were conducted on QUBO instances of size $n$ = 1,000 and $n$ = 6,000 across all threshold values. Each configuration was executed five times using different initial random seeds to account for variations introduced by stochastic initialization. The number of optimization steps, total runtime, and final solution energy were recorded for every run to support comparative analysis with the other solvers.

**Verification of Solution Energies.** To ensure the correctness and consistency of the solutions obtained from different QUBO solvers, we implemented a verification script that independently recalculates the energy value of each binary solution vector. All verification routines were executed on the CPU using PyTorch, providing consistent numerical behavior across all test cases.

This verification step plays a critical role in benchmarking, debugging, and ensuring the reproducibility of results in QUBO-based optimization research. It ensures that each reported solution is structurally valid and energetically accurate concerning the original QUBO objective function.

**Experimental Platform.** All experiments were conducted on a Linux workstation running Ubuntu 22.04. The system was equipped with an Intel Core i7-14700 processor (20 physical cores, 28 threads), 62 GB of RAM, and an NVIDIA GeForce RTX 4070 SUPER GPU with 16 GB of VRAM, utilizing CUDA version 12.2.

The software environment was based on Python 3.11.9. Solver implementations were developed using the following libraries: PyTorch 2.4.1, JAX 0.5.3, SciPy 1.14.1, Optax 0.2.4, NumPy 1.26.4, and PyQUBO 1.5.0. All experiments were performed within an isolated Conda virtual environment to ensure reproducibility and environment consistency. GPU-based solvers leveraged CUDA acceleration, while CPU-based methods utilized all available logical cores through multi-threaded execution.

## Results

**Energy Minimization Results on $Q \in \mathbb{R}^{1,000 \times 1,000}$.** Figure 1 presents the energy values obtained from five QUBO solvers: Neal, PyTorch (CPU), PyTorch (GPU), JAX, and SciPy, across five independently generated QUBO matrices of size 1,000×1,000. The x-axis represents the convergence threshold, ranging logarithmically from $10^{-1}$ to $10^{-6}$, while the y-axis shows the final objective value of the QUBO function after optimization.

In all five test cases (a) to (e), Neal consistently produced low energy values. Both PyTorch (CPU) and PyTorch (GPU) achieved comparable results, with energy values gradually decreasing as the threshold became more stringent. JAX exhibited moderate improvements in energy with decreasing thresholds. SciPy showed the most pronounced dependence on the convergence threshold, yielding significantly higher energy values at loose thresholds and visibly improving as the threshold tightened. For most solvers, energy values converged by the threshold reached $10^{-4}$, indicating diminishing returns from further tightening.

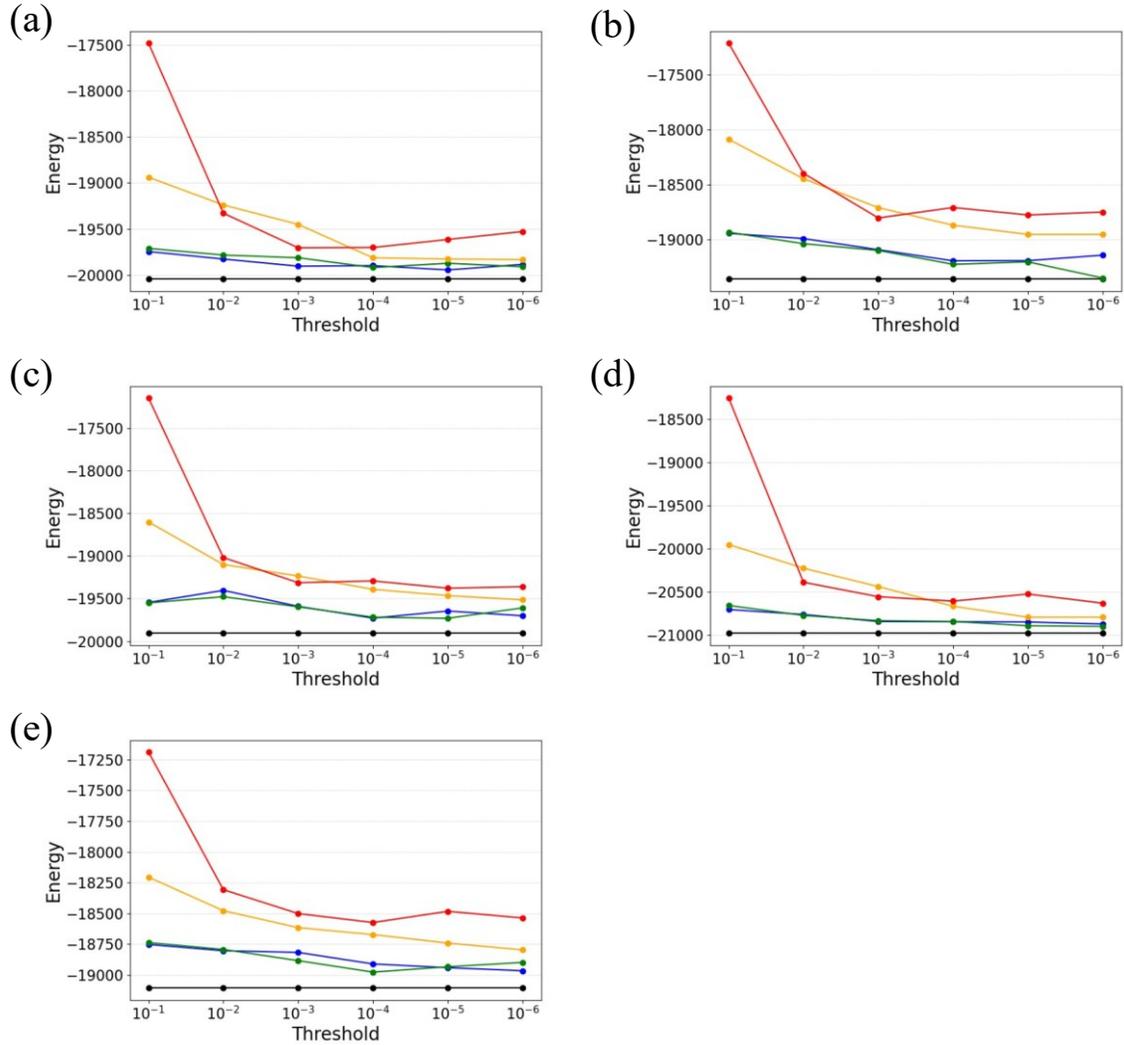

**Figure 1.** Energy values were obtained from five QUBO solvers under varying convergence thresholds. Each subfigure (a) through (e) corresponds to a distinct randomly generated QUBO matrix $Q \in \mathbb{R}^{1,000 \times 1,000}$. The x-axis represents the stopping threshold applied during optimization, ranging from $10^{-1}$ to $10^{-6}$, and the y-axis shows the final QUBO objective value (Energy), where lower values indicate better solution quality. Solver results are color-coded: black for Neal, blue for PyTorch (CPU), green for PyTorch (GPU), orange for JAX, and red for SciPy.

**Energy Minimization Results on $Q \in \mathbb{R}^{6,000 \times 6,000}$.** Figure 2 illustrates the energy values obtained by the five solvers on five independently generated QUBO matrices of size 6,000×6,000, with threshold values ranging from $10^{-1}$ to $10^{-6}$. The results follow a similar pattern to those observed for the more minor 1,000×1,000 cases. In all five subfigures (a) through (e), PyTorch (CPU), PyTorch (GPU), and JAX consistently achieved low energy values that improved as the threshold decreased. Neal also produced stable and competitive results. In contrast, SciPy maintained significantly higher energy values across all thresholds and exhibited minimal improvement under stricter convergence conditions. As the problem size increased to 6,000 variables, the performance gap between SciPy and the other solvers became more pronounced. At the same time, Neal, PyTorch, and JAX continued to produce low and stable energy values as the threshold tightened.

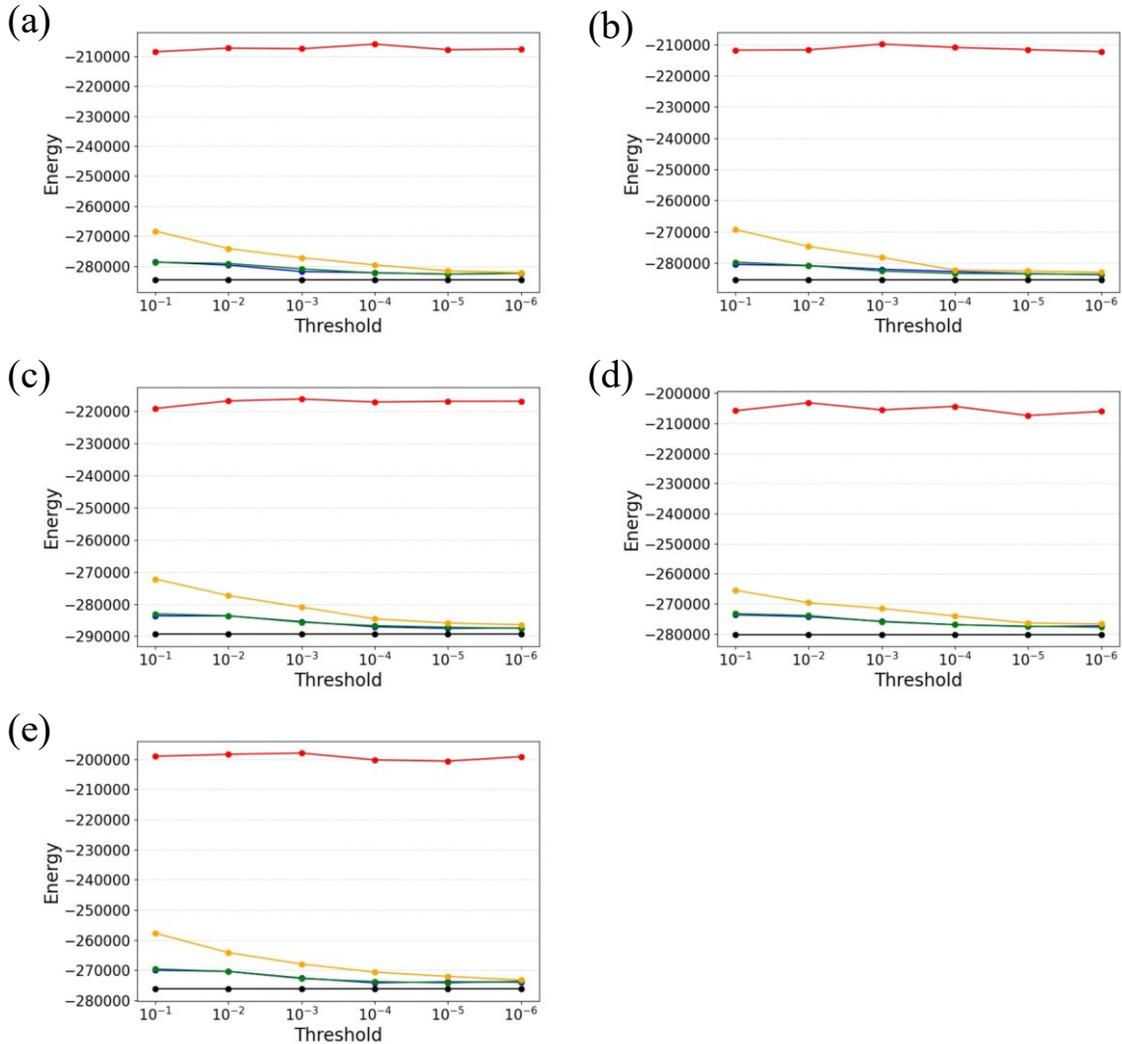

**Figure 2.** It is the same setup as in Figure 1 but applied to QUBO matrices $Q \in \mathbb{R}^{6,000 \times 6,000}$. Each subfigure (a) through (e) corresponds to a distinct randomly generated matrix. The x-axis represents the convergence threshold from $10^{-1}$ to $10^{-6}$, and the y-axis shows the resulting QUBO energy after optimization. Color representations for each solver are consistent with Figure 1.

**Energy Minimization Results on $Q \in \mathbb{R}^{25,000 \times 25,000}$.** Figure 3 shows the energy values obtained from PyTorch (CPU), PyTorch (GPU), and JAX on five randomly generated QUBO matrices of size 25,000×25,000, evaluated under six convergence thresholds ranging from $10^{-1}$ to $10^{-6}$. Due to memory limitations, Neal and SciPy were excluded from this experiment, as their memory requirements were significantly higher and unsuitable for problems of this scale.

Across all five subfigures (a) through (e), the three included solvers consistently demonstrated improved energy values as the threshold decreased. JAX exhibited the most incredible sensitivity to convergence criteria, achieving substantial energy reduction with tighter thresholds. PyTorch (CPU) and PyTorch (GPU) produced similarly low energy values, with moderate gains observed under stricter settings. Compared to experiments on smaller matrix sizes, the energy values in this case were significantly larger, reflecting the increased dimensionality of the optimization problem. The performance differences among the solvers became more pronounced at this scale.

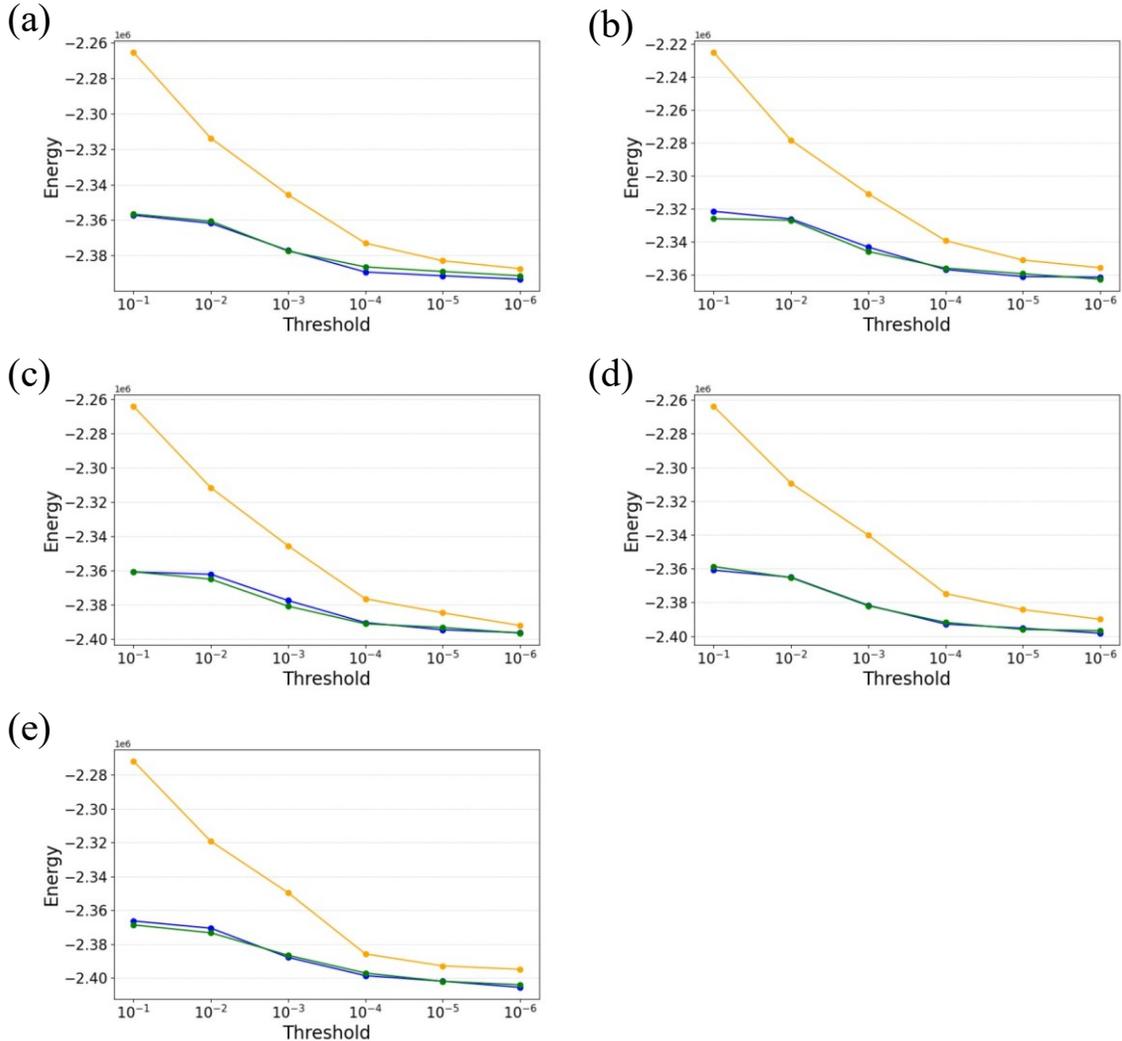

**Figure 3.** It is the same experimental setup as in Figure 1 but applied to QUBO matrices $Q \in \mathbb{R}^{25,000 \times 25,000}$. Each subfigure (a) through (e) corresponds to a distinct randomly generated matrix. The x-axis shows the convergence threshold values ranging from $10^{-1}$ to $10^{-6}$, while the y-axis indicates the final energy obtained after optimization. Solver color codes are consistent with previous figures.

**Energy Minimization Results on $Q \in \mathbb{R}^{45,000 \times 45,000}$.** Figure 4 presents the energy values obtained using PyTorch (GPU) on five randomly generated QUBO matrices of size 45,000×45,000, evaluated under six convergence thresholds ranging from $10^{-1}$ to $10^{-6}$. Due to memory constraints, only PyTorch (GPU) was used in this large-scale experiment. Other solvers, including Neal, JAX, and SciPy, were excluded as their memory requirements exceeded the available resources for problems of this size.

In all five subfigures (a) through (e), energy values consistently decreased as the convergence threshold became more stringent. This trend indicates that the optimization process continued to improve solution quality under tighter stopping criteria, even at this high-dimensional scale. Despite the computational intensity, PyTorch (GPU) maintained numerical stability and demonstrated scalable performance, producing reliable reductions in the objective value across all convergence settings.

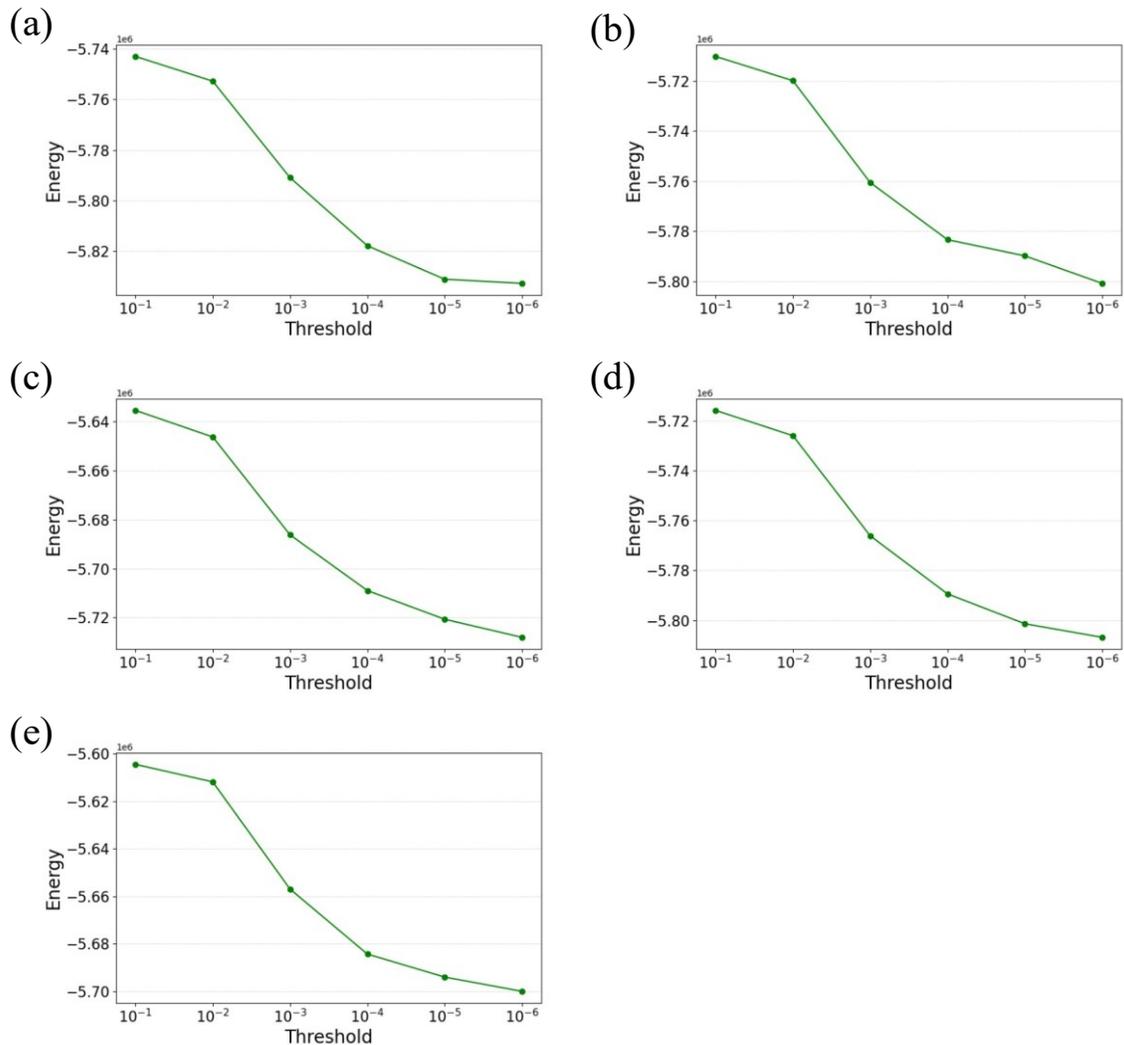

**Figure 4.** Same experimental setup as in Figure 1, but applied to QUBO matrices $Q \in \mathbb{R}^{45,000 \times 45,000}$. Each subfigure (a) through (e) corresponds to a distinct randomly generated

matrix. The x-axis represents convergence thresholds ranging from $10^{-1}$ to $10^{-6}$, and the y-axis shows the final QUBO energy values. Only PyTorch (GPU) was evaluated in this large-scale setting due to memory and runtime constraints.

**Computation Time Analysis across Solvers and Problem Sizes.** Figure 5 summarizes the computation time each solver requires across different convergence thresholds for progressively larger QUBO matrix sizes. Subfigures (a) through (d) correspond to problem sizes of 1,000×1,000, 6,000×6,000, 25,000×25,000, and 45,000×45,000, respectively. The x-axis indicates the stopping threshold, ranging from $10^{-1}$ to $10^{-6}$, while the y-axis represents runtime in seconds.

All solvers were completed within a few seconds for the smallest problem size (Figure 5a). SciPy and JAX showed increasing runtime as the threshold became more stringent, while PyTorch (CPU and GPU) maintained relatively stable computation times. As the problem size increased to 6,000×6,000 (Figure 5b), overall runtimes grew noticeably, with SciPy remaining the slowest solver.

At the 25,000×25,000 scale (Figure 5c), only PyTorch (CPU), PyTorch (GPU), and JAX were evaluated due to memory limitations. Among them, PyTorch (CPU) required the most extended runtime, which increased sharply at stricter thresholds. JAX and PyTorch (GPU) demonstrated better scalability, with PyTorch (GPU) achieving the shortest runtime in large-scale cases.

For the enormous problem size, 45,000×45,000 (Figure 5d), only PyTorch (GPU) completes all runs. The computation time increased steadily as the convergence threshold became more stringent, reaching over 2,500 seconds at the highest precision level. Although PyTorch (CPU) can also execute this problem size on the current hardware setup, it requires significantly more time to complete. Therefore, CPU-based runs were not included in this experiment.

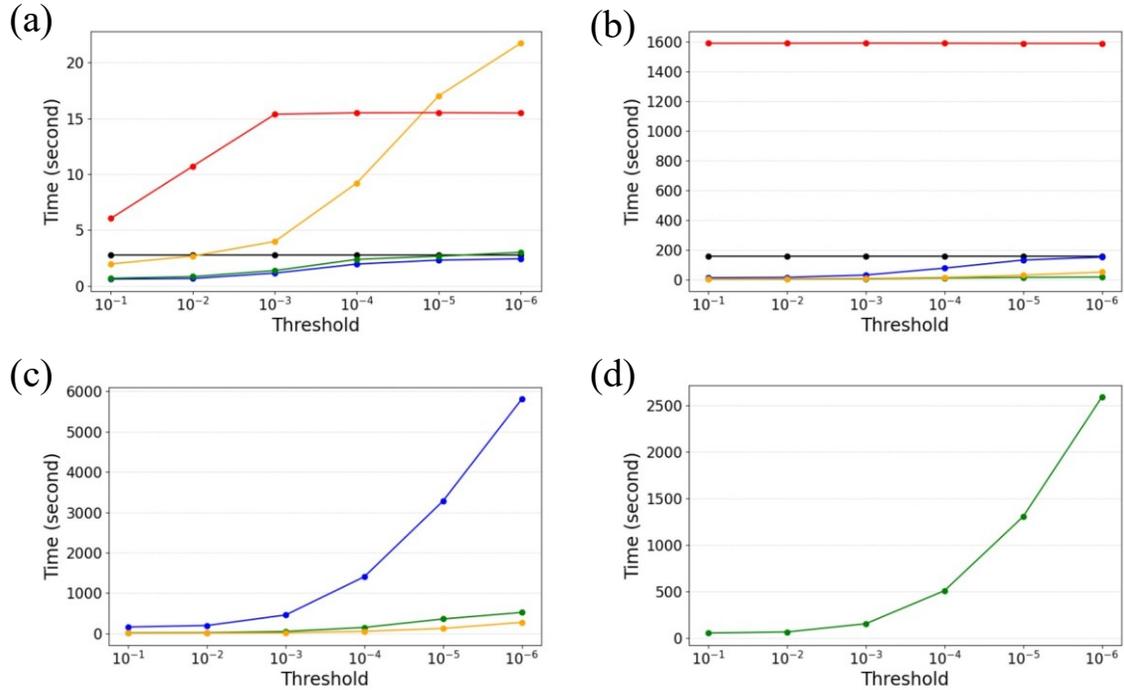

**Figure 5.** The computation time required by different QUBO solvers under varying convergence thresholds. Subfigures (a) through (d) correspond to QUBO matrices of increasing size: (a) 1,000×1,000, (b) 6,000×6,000, (c) 25,000×25,000, and (d) 45,000×45,000. The x-axis shows the stopping threshold, ranging from $10^{-1}$ to $10^{-6}$, and the y-axis indicates runtime in seconds. Solver color codes are as follows: black for Neal, blue for PyTorch (CPU), green for PyTorch (GPU), orange for JAX, and red for SciPy. As matrix size and convergence precision increase, CPU-based and SciPy solvers exhibit substantial growth in computation time, whereas GPU-based solvers remain more scalable and time-efficient.

**Discussion**

The experimental results presented in Figures 1 through 5 reveal consistent patterns regarding the performance, scalability, and computational behavior of five QUBO solvers across varying problem sizes and convergence thresholds. Each solver exhibited distinct strengths and limitations, especially as the QUBO matrix scale increased from 1,000×1,000 to 45,000×45,000.

Regarding solution quality, PyTorch (CPU and GPU), JAX, and SciPy all demonstrated improvement in energy minimization as the convergence threshold decreased from $10^{-1}$ to $10^{-6}$, as shown in Figures 1 to 4. This trend suggests that stricter convergence criteria generally yield better optimization outcomes. PyTorch consistently produced high-quality solutions, even under relaxed thresholds, making it a robust choice across all problem sizes. JAX was also competitive, particularly for small- to mid-scale matrices. SciPy, by contrast, showed strong sensitivity to threshold settings, performing poorly under loose criteria and gradually improving only under more stringent ones. Due to memory limitations, Neal maintained low and stable energy values for small-scale problems but was not tested on larger matrices.

Scalability became a critical issue as matrix size increased. Only memory-efficient solvers such as PyTorch and JAX remained viable at larger scales. In particular, Figures 3 and 4 indicate that only PyTorch (GPU) could handle the enormous problem size of 45,000×45,000 without exhausting system resources. JAX reached its practical limit at 25,000×25,000, while SciPy and Neal were constrained to more minor problems due to their high memory overhead. These results highlight the importance of considering solver memory consumption when tackling significant combinatorial optimization problems.

Regarding runtime, as depicted in Figure 5, PyTorch (GPU) displayed the best scalability across increasing thresholds and matrix sizes. It maintained reasonable wall-clock times even for the largest matrices. PyTorch (CPU), though slower, still completed all runs and showed stable performance relative to problem complexity. JAX exhibited efficient runtime at moderate scales but was excluded from the most extensive matrix due to resource constraints. SciPy consistently had the longest execution times, especially as the threshold tightened, and its runtime appeared to plateau due to internal algorithmic constraints. Neal remained fast in small instances but was not tested at larger scales.

In summary, PyTorch, particularly its GPU-accelerated implementation, emerged as the most balanced solver regarding energy quality, runtime, and scalability. JAX offers competitive performance at moderate scales with lower memory usage, while SciPy and Neal, though effective in minor cases, are limited by memory requirements. These findings underscore the importance of selecting solvers based on expected optimization accuracy, available hardware resources, and the problem's dimensionality.

**Conclusion**

This study systematically evaluated the performance of five software-based QUBO solvers, Neal, PyTorch (CPU), PyTorch (GPU), JAX, and SciPy, across multiple problem sizes and convergence thresholds. The results demonstrate that solution quality, runtime efficiency, and scalability vary considerably depending on the solver architecture and parameter settings.

PyTorch, particularly with GPU acceleration, consistently produced high-quality solutions and exhibited strong scalability across problem sizes up to 45,000×45,000, making it a robust and versatile option for large-scale QUBO problems. JAX also showed competitive performance in energy minimization and runtime at moderate scales, although it encountered memory limitations at the largest problem size. SciPy was highly sensitive to the convergence threshold, often requiring longer runtimes and producing lower-quality solutions under loose stopping conditions. Neal demonstrated fast and stable performance for small to mid-sized problems but could not scale to larger dimensions due to memory constraints.

These findings underscore the importance of selecting QUBO solvers that balance convergence accuracy, memory consumption, and computational efficiency according to the problem's specific scale and the hardware resources available.

**Competing interests**

The authors declare no competing interests.

**Data and Software Availability**

The data supporting the findings of this study are openly available on GitHub at the following URL: https://github.com/peikunyang/11_Qubo_solver.